\documentclass{PoS}

\usepackage{heppennames2}

\newcommand{\ttbar}{\ensuremath{\PQt\PAQt}\xspace} 

\newcommand{\tcg}{\ensuremath{\HepProcess{\PQt \to \PQc \PGg}}\xspace}
\newcommand{\tcx}{\ensuremath{\HepProcess{\PQt \to \PQc \HepParticle{E}\!\!\!\!\!\!\!\slash}}\xspace}

\graphicspath{{./plots/}}

\title{News on the CLIC physics potential}

\ShortTitle{News on the CLIC physics potential}

\author{
  \speaker{Aleksander Filip \.Zarnecki}
  ~~~  
        on behalf of the CLICdp Collaboration \\
        Faculty of Physics, University of Warsaw\\
        E-mail: \email{Filip.Zarnecki@fuw.edu.pl}
}

\abstract{
The Compact Linear Collider (CLIC) is a proposed TeV-scale
high-luminosity electron-positron collider.
For an optimal exploitation of its physics potential, CLIC is foreseen
to be built and operated in three stages, with centre-of-mass energies
ranging from 380\,GeV up to 3\,TeV.
Electron beam polarisation is provided at all energies.
The initial energy stage will focus on precision measurements of
Higgs-boson and top-quark properties.
The subsequent energy stages enhance the reach of many direct and
indirect searches for new physics Beyond the Standard Model (BSM) and give
access to the Higgs self-coupling.
Higgs and top-quark projections have been evaluated using full
detector simulation studies.
Many new phenomenology studies have been undertaken to explore the BSM
reach of CLIC, from Effective Field Theory (EFT) interpretations of
precision measurements through to signature-based searches; these
include flavour dynamics, and dark matter and heavy neutrino searches. 
Selected results that demonstrate the outstanding potential of CLIC in
many physics domains are reviewed.
}

\FullConference{ALPS 2019 An Alpine LHC Physics Summit \\
		April 22 - 27, 2019\\
		Obergurgl, Austria}

\begin{document}

\section{Introduction}

The Compact Linear Collider (CLIC) is a mature option for a future
electron-positron collider operating at centre-of-mass energies of
up to 3\,TeV.
The conceptual design for a machine based on a two-beam acceleration
scheme, allowing for accelerating gradients up to 100\,MV/m, was
presented in 2012\,\cite{Aicheler:2012bya}.  
For full exploitation of its physics potential
CLIC will be built and operated in three stages\,\cite{Charles:2018vfv},
with accelerator footprint of 11 to 50 km.
The energy of 380\,GeV was chosen for the initial stage to optimize the 
physics potential in terms of Higgs-boson and top-quark measurements. 
Assuming 8 years of running at this stage, a total of
1\,ab$^{-1}$ of data should be collected, including  100\,fb$^{-1}$ at
the \ttbar threshold.
This is similar to the luminosity expected per interaction point for
\mbox{FCC-ee}\,\cite{Abada:2019zxq} with half the construction costs and half
the power consumption of the initial stage of CLIC.
Two high energy stages of CLIC, at $\sqrt{s} = 1.5$ TeV with an assumed
integrated luminosity of 2.5\,ab$^{-1}$ and at $\sqrt{s} = 3$ TeV with
5\,ab$^{-1}$, will mainly focus on direct and indirect Beyond the
Standard Model (BSM) searches, but additional Higgs boson and
top-quark studies are also possible. 
Some of the rare Standard Model processes can only be measured at the
TeV energies.
CLIC is the only e$^+$e$^-$ collider project that can go into this domain.

A dedicated CLIC detector concept, CLICdet, has been developed
to assure full exploitation of physics potential from 380\,GeV to 3\,TeV,
based on detailed simulation studies, detector R\&D programme and beam
tests\,\cite{Arominski:2018uuz}.
The design is optimised for Particle Flow
reconstruction\,\cite{Thomson:2009rp} with expected 
jet energy resolution  $\sigma_{E}/{E} = 3 - 4$\% for high jet energies.
High resolution tracking based on all-silicon vertex and tracker detectors
will also allow for very efficient flavour tagging.

Presented in the contribution are selected results based on
detector-level simulations from recent 
studies of the physics potential of CLIC.
For more information on the CLIC project and its physics potential,
please refer to the documents submitted as input to the
European Strategy for Particle Physics Update 2018-2020\,\cite{esu}.

\section{Higgs physics}

\begin{figure}[tb]
\begin{minipage}{0.48\linewidth}\centering
  \includegraphics[width=\textwidth]{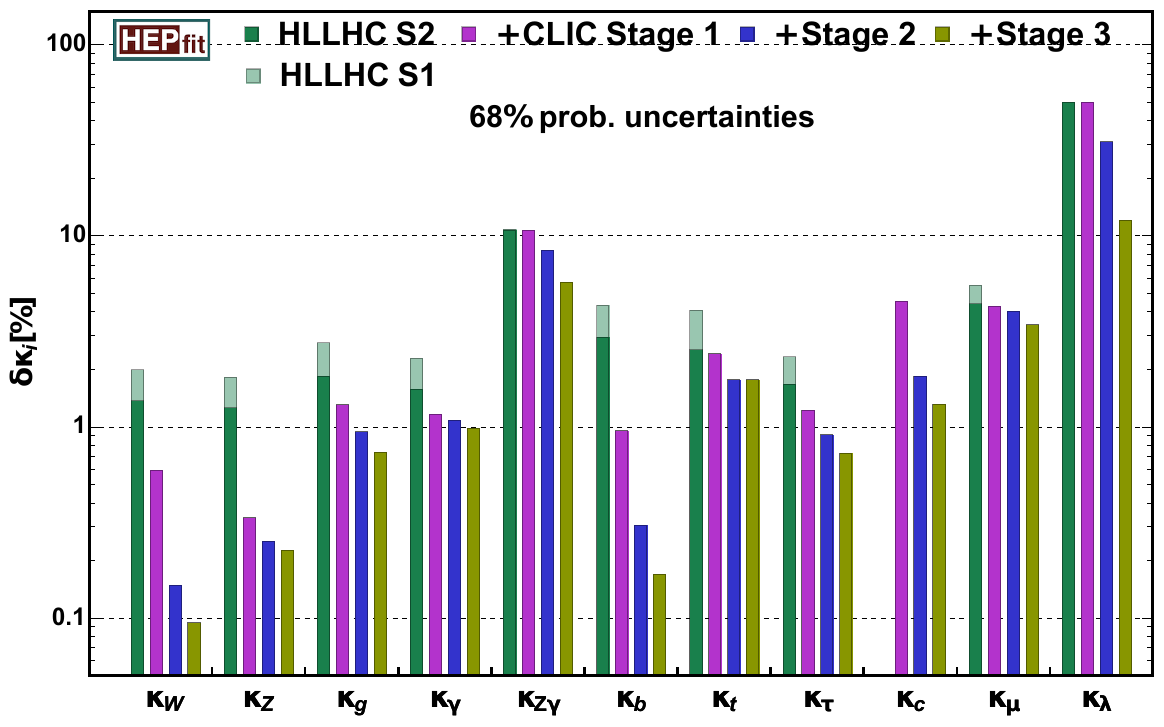}
  ~
\end{minipage}
\begin{minipage}{0.51\linewidth}\centering
\includegraphics[width=\textwidth]{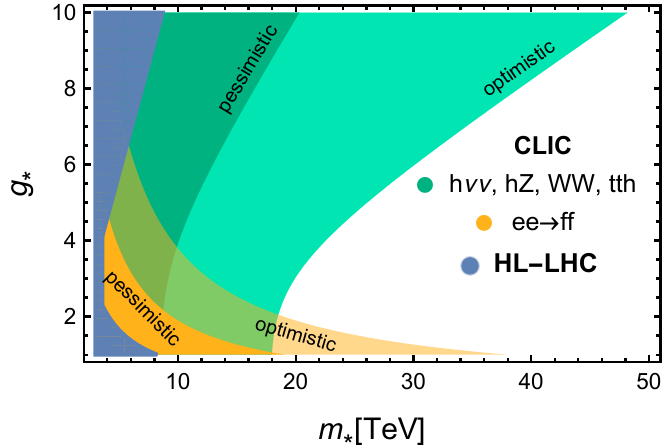}
\end{minipage}
\caption{
Left: CLIC sensitivity to the different Higgs boson couplings from the
model-dependent fit (combined with the HL-LHC projections).
Right: discovery (5$\sigma$) reach on Composite Higgs at CLIC,
compared with expected exclusion limits (2$\sigma$) from the HL-LHC.
Figures taken from\,\cite{deBlas:2018mhx}.
}\label{fig:higgs}
\end{figure}

Two main channels contribute to the Higgs boson production
at CLIC: the Higgsstrahlung process, $\Pe^+ \Pe^- \to \PZ \PH$, and the
WW-fusion process,  $\Pe^+ \Pe^- \to \PH \PGn \PAGn$.
Recoil mass reconstruction in Higgsstrahlung events allows for
unbiased event selection, which is crucial for model-independent
determination of the Higgs couplings\,\cite{Abramowicz:2016zbo}.
The clean environment allows for unambiguous separation of different decay
channels opening prospects for direct measurement of BR($\PH \to \PQc
\PAQc$) and setting stringent limits on the invisible Higgs boson decays.
By combining all cross section measurements for the two production
channels, the Higgs boson couplings can be extracted with a
model-independent fit.
For example, the coupling to the $\PZ$ boson can be established with 
a precision of 0.6\% at the first energy stage alone\,\cite{Robson:2018zje}. 
When including data collected at 1.5\,TeV and 3\,TeV, most of the
Higgs couplings can be measured at the percent level and the total
width of the boson can be determined to 2.5\%.
In Fig.~\ref{fig:higgs}\,(left)
CLIC sensitivity to the different Higgs boson couplings 
is compared with the HL-LHC projections, for a model-dependent
analysis\,\cite{deBlas:2018mhx}. 
For some of the couplings, CLIC measurements will reduce their
uncertainties by an order of magnitude.
High energy stages of CLIC allow also the Higgs-boson pair
production process to be studied. 
The trilinear Higgs self-coupling, which is crucial for understanding of
electroweak symmetry breaking,  can be extracted from the measured
kinematic distributions with uncertainty of the order of
10\%\,\cite{Roloff:2019crr}. 
Precision  measurements at CLIC can be also used to search for or constrain
different new physics scenarios.
Shown in  Fig.~\ref{fig:higgs}\,(right) is the 5$\sigma$ discovery range 
for Higgs compositeness at CLIC compared to expected HL-LHC 2$\sigma$
exclusions limits\,\cite{deBlas:2018mhx}. 
For this type of model, the discovery reach of CLIC can extend beyond the
HL-LHC energy scale.
 
\section{Top quark physics}

\begin{figure}[tb]
\begin{minipage}{0.4\linewidth}\centering
\includegraphics[width=\textwidth]{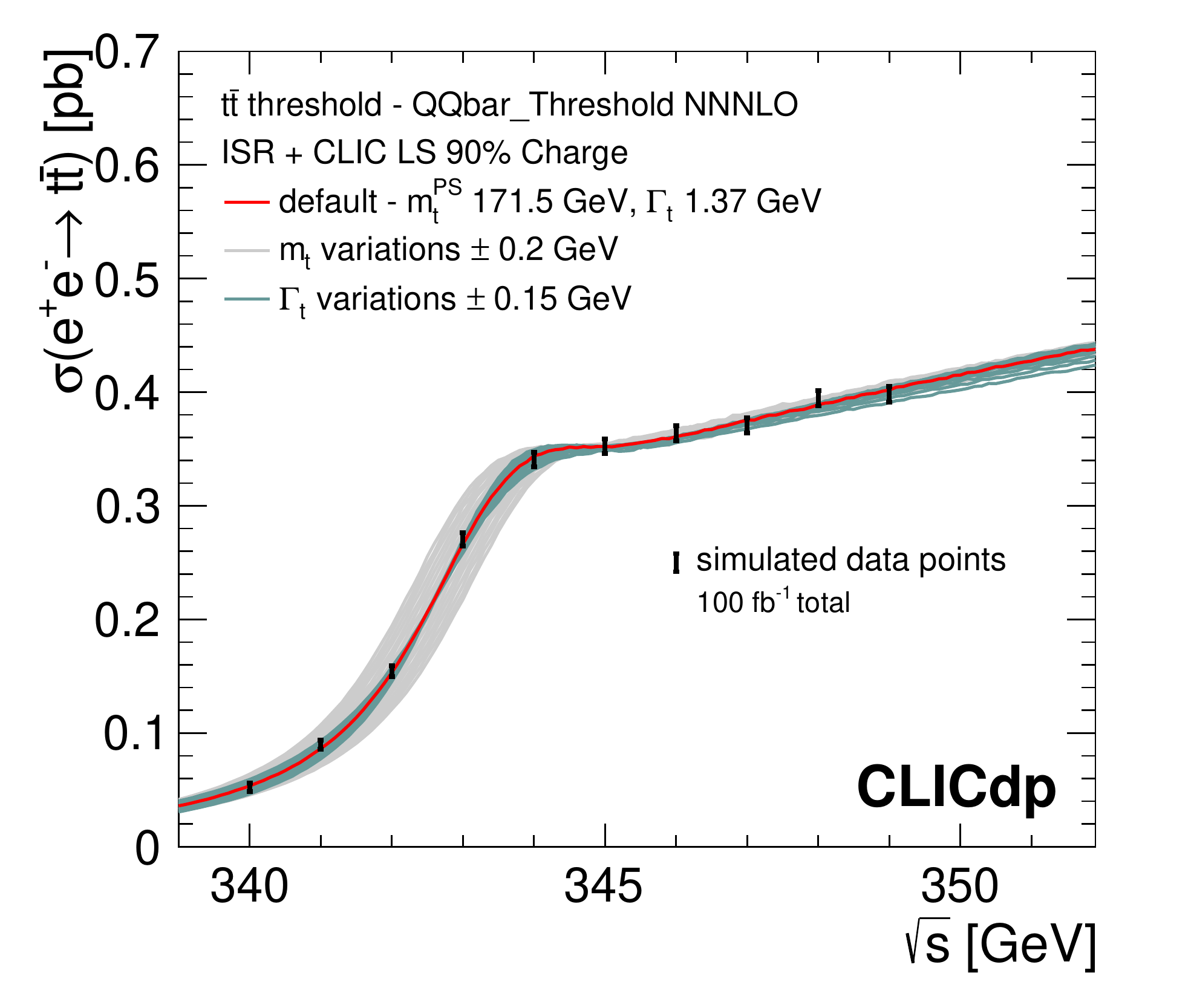}
\end{minipage}
\begin{minipage}{0.6\linewidth}\centering
\includegraphics[width=\textwidth]{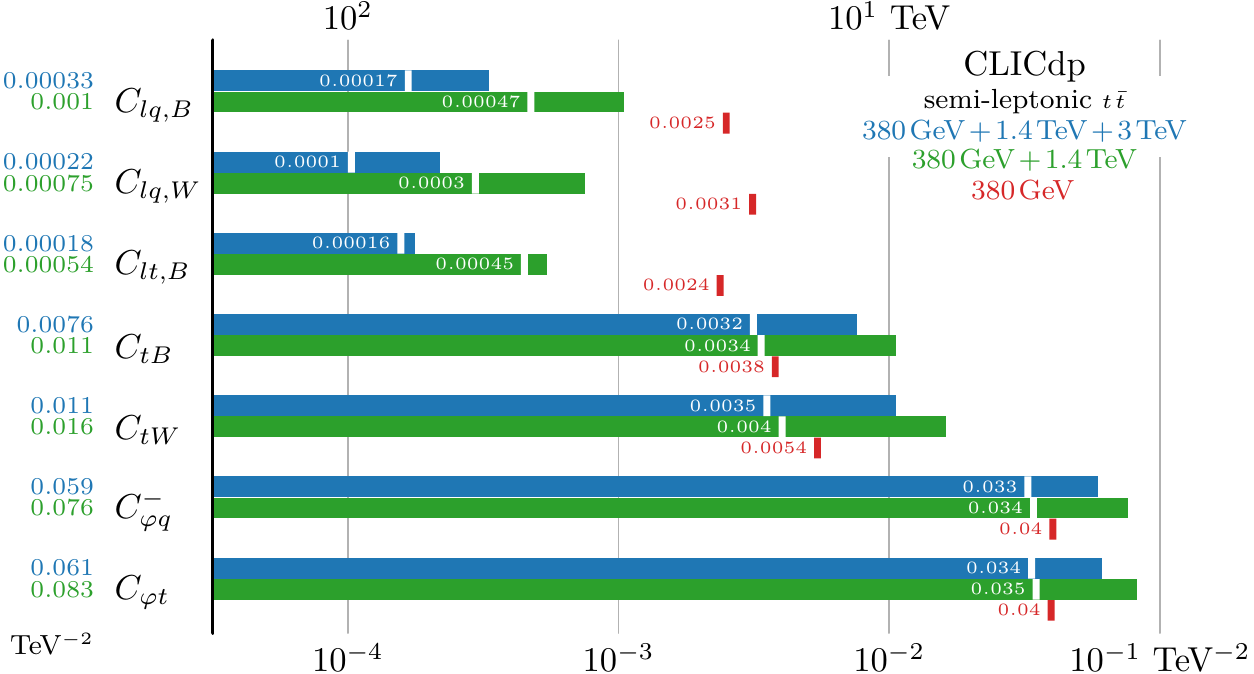}
\end{minipage}
\caption{
Left: top-quark mass determination from the dedicated threshold scan
at CLIC.
Right: summary of the global EFT analysis results using statistically
optimal observables for the three CLIC energy stages.
Figures taken from \cite{Abramowicz:2018rjq}.
}\label{fig:top}
\end{figure}

Already at the initial stage of CLIC over a million top quarks and anti-quarks
will be produced.
This will allow for detailed study of top-quark properties including 
top-quark mass and electroweak couplings, and searching for rare top-quark
decays.
For a direct mass measurement from reconstruction of top-quark
hadronic decays a statistical precision of 30\,MeV is expected.
However, the measurement requires excellent control of the jet energy
scale and is subject to substantial theoretical uncertainties.
The most precise determination of the top-quark mass, both in terms of the
statistical and of the systematic uncertainties, is possible with a
dedicated scan of the top pair production threshold.
With 100\,fb$^{-1}$ collected in 10 steps around the threshold (see
Fig.~\ref{fig:top} left), a statistical mass
uncertainty of 20--30\,MeV is expected and a total systematic
uncertainty of the order of 50\,MeV\,\cite{Abramowicz:2018rjq}.

With large statistics of top quarks and anti-quarks produced, rare
top-quark decays can be searched for. Expected limits for
Flavour-Changing Neutral Current (FCNC) decays involving the $\PQc$ quark,
assuming 1000\,fb$^{-1}$ collected at 380\,GeV, range from $2.6 \times
10^{-5}$ for BR(\tcg) to $3.4 \times 10^{-4}$ for a decay involving
an invisible dark matter particle, BR(\tcx).
FCNC top couplings can also be constrained from the measurement of
single top production. 
These constraints are comparable to the direct ones for 380\,GeV,
but improve significantly  for higher energy running\,\cite{deBlas:2018mhx}.

The top-quark Yukawa coupling can be  indirectly constrained from the
threshold scan.
However, the measurement is very sensitive to systematic effects.
Direct determination of the top-quark Yukawa coupling is possible from
the measurement of the $\PQt\PAQt\PH$ production cross section at high
energy stages of CLIC.
Signal event selection, with four b-jets expected in the final state,
profits from the excellent flavour tagging capabilities.
The expected precision on the top Yukawa coupling  for 2.5\,ab$^{-1}$
of data collected at 1.5\,TeV is 2.7\%.

Precise measurements of the top-quark pair-production cross-sections,
forward-backward asymmetries and angle distributions in top decays
can be use to set constraints on possible BSM contributions in the
framework of Effective Field Theory (EFT) operators. 
Measurements at one energy stage are insufficient to simultaneously
constrain all relevant EFT couplings.
Only by combining data collected at different energies (and polarisations)
all Wilson coefficients contributing to the top-quark pair production
can be constrained simultaneously. 
Summary of the global EFT analysis results, based on statistically
optimal observables, is shown in Fig.~\ref{fig:top}\,(right).
With high precision of top measurements CLIC is sensitive to ``new physics''
in the top sector up to scales in the 100\,TeV range.

\section{Beyond Standard Model physics}

\begin{figure}[tb]\centering
\includegraphics[width=0.95\textwidth]{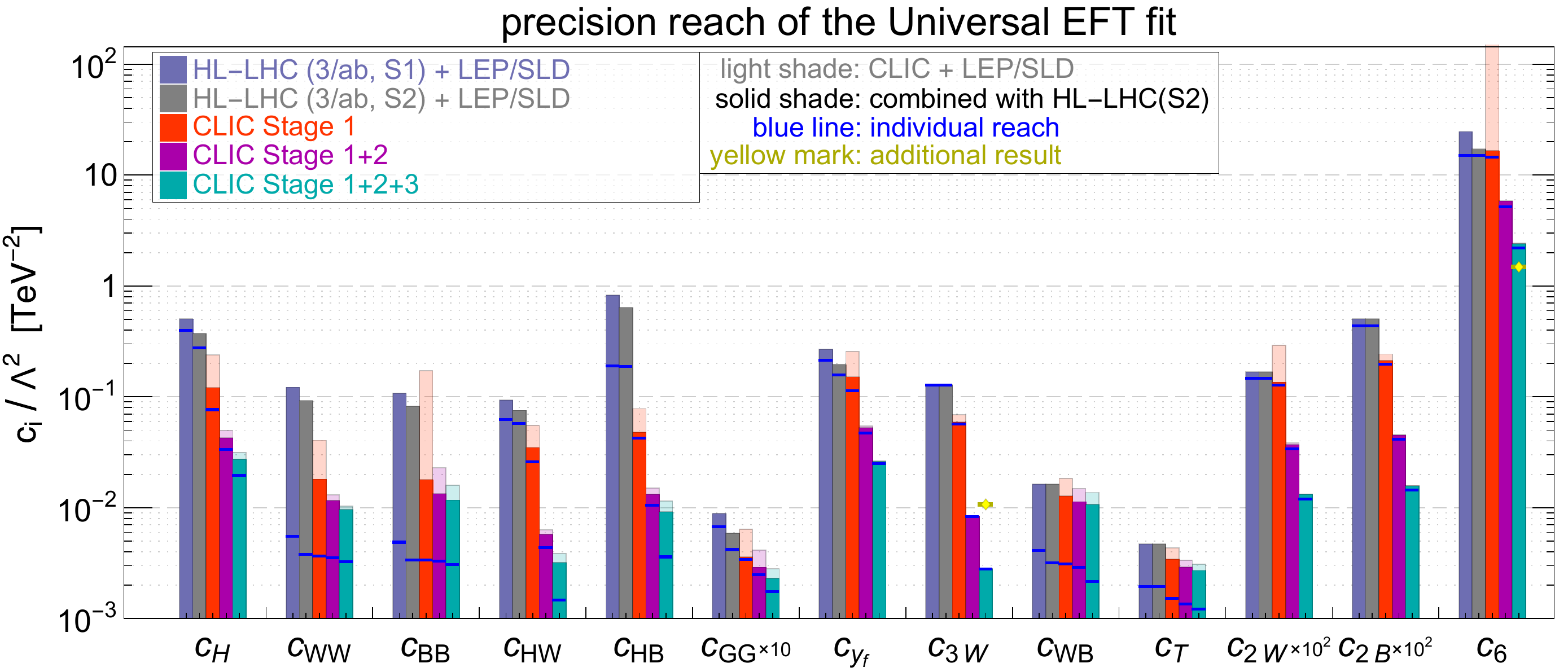}
\caption{
 Summary of the CLIC sensitivity to EFT operators $c_i /\Lambda^2$ from
 a global analysis of Higgs and top-quark observables, WW production,
 and two-fermion scattering processes, for three energy stages.
 Figure taken from \cite{deBlas:2018mhx}.
}\label{fig:eft}
\end{figure}

While strong limits on mass scales of ``new physics'' are expected at
HL-LHC for scenarios involving strongly interacting new particles,
complementary searches are possible at CLIC.
Both direct searches, for models with weak couplings or soft signatures,
as well as high sensitivity indirect searches, based on precision
measurements, were considered\,\cite{deBlas:2018mhx}.
Sensitivity to possible ``new physics'' processes at large energy
scales can be studied in the general framework of the EFT operators. 
By combining precision measurements of different Higgs and top-quark
observables, as well as  WW  and two-fermion production processes,
limits on the considered EFT operators can be set in the global analysis.
As shown in Fig.~\ref{fig:eft}, CLIC measurements will significantly
improve limits expected from HL-LHC.

\begin{figure}[tb]
\begin{minipage}{0.5\linewidth}\centering
\includegraphics[width=\textwidth]{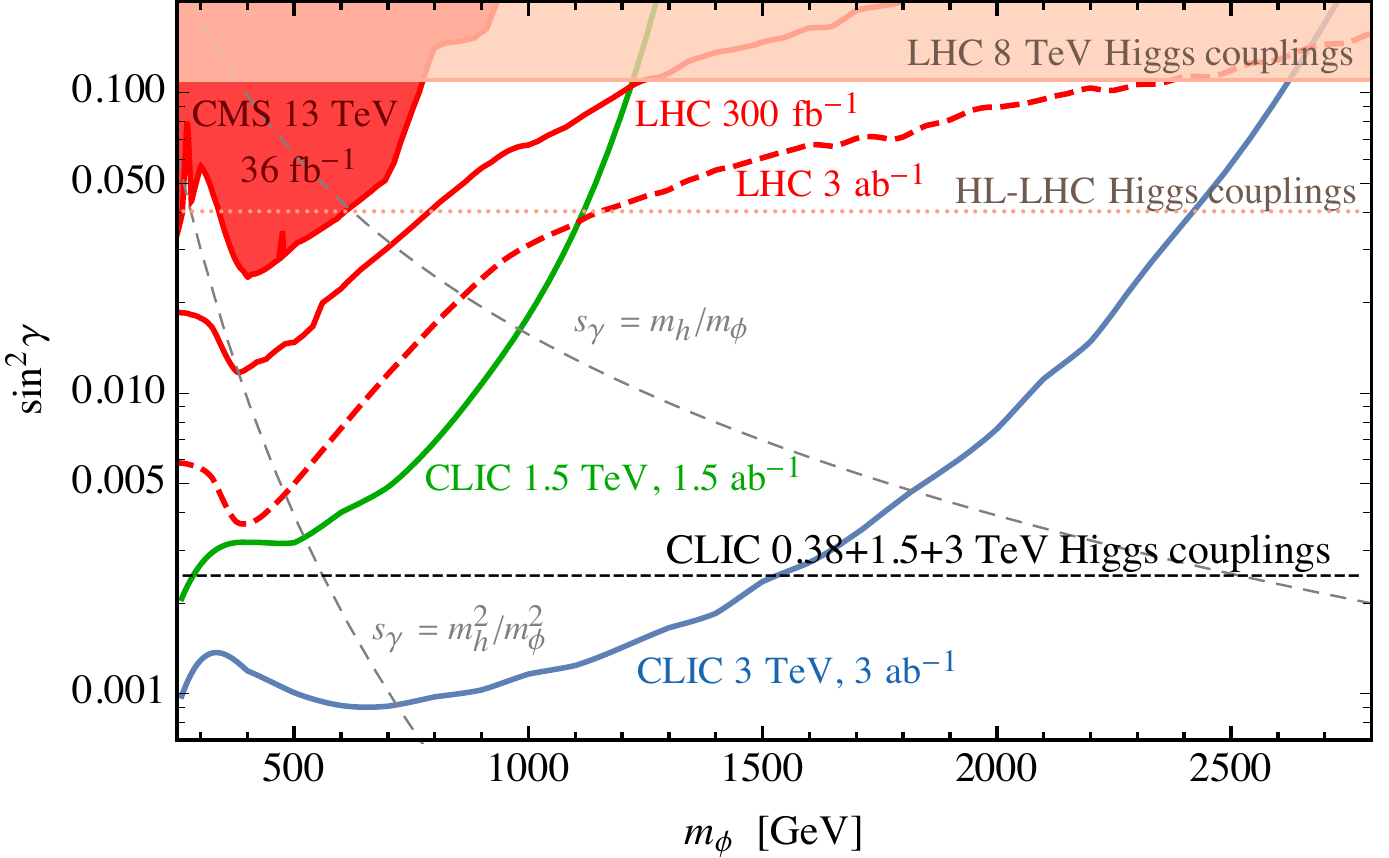}
\end{minipage}
\begin{minipage}{0.5\linewidth}\centering
\includegraphics[width=\textwidth]{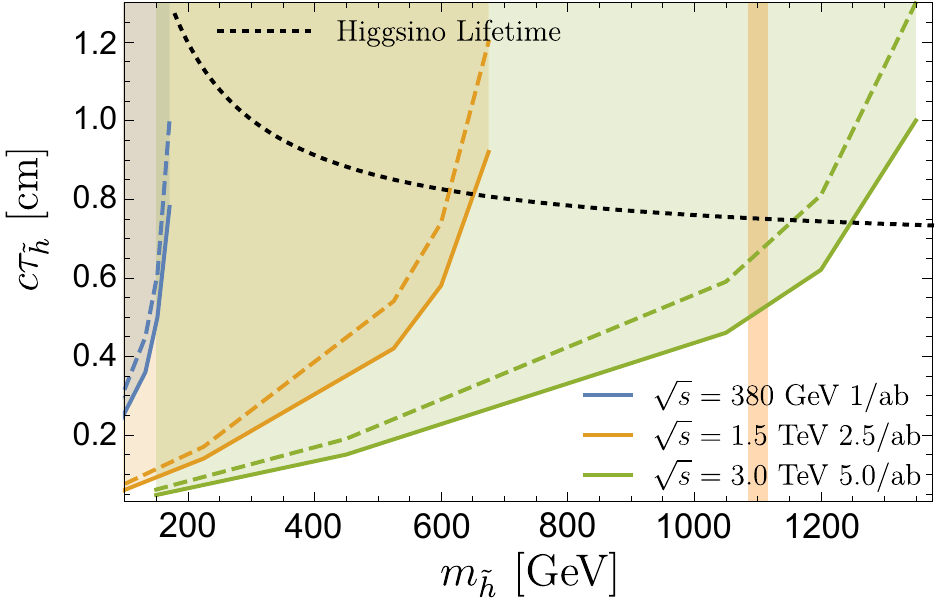}
\end{minipage}
\caption{
  Left: expected 95\% C.L. exclusion limits from direct search for heavy
  scalar singlet production at CLIC compared with the expected limits from
  LHC and HL-LHC and the expected indirect limits from Higgs boson
  coupling measurements.
  Right: 95\% C.L. exclusion contours in lifetime-mass for N = 3
  (solid) and N = 30 (dashed) Higgsino events in the detector
  acceptance at the three stages of CLIC. 
  Figures taken from \cite{Charles:2018vfv}.
}\label{fig:bsm}
\end{figure}

For many models, in particular those with an exotic scalar sector
or new Higgs bosons, the CLIC direct and indirect reach can exceed
that of HL-LHC.
As an example, indirect and direct sensitivities to new heavy scalar singlets
are shown in Fig.~\ref{fig:bsm}\,(left).
For the direct search, the heavy scalar decay to a pair of Higgs bosons,
$\PGF \to \PH\PH \to \PQb\PAQb\PQb\PAQb$, was considered.

The Higgsino is one of the common targets in searches for
supersymmetric extensions of the Standard Model.
If it is not the lightest supersymmetric particle, the Higgsino can decay
into a SM boson and missing energy.
One of the possible signatures are the ``disappearing tracks'' caused by
a heavy charged particle passing through the detector before decaying
to a neutral particle that escapes detection and a very soft charged
particle, unresolved in the detector.
Search for dark matter production using ``disappearing tracks''
signature at CLIC profit from  precision tracking and low
background conditions.
Expected exclusion contours for the three stages of CLIC are compared
in Fig.~\ref{fig:bsm}\,(right).

\section{Conclusions}

CLIC is an attractive and cost-effective option for a next large
facility at CERN offering the unique combination of high collision
energies and the clean environment of electron-positron collisions.
The energy chosen for the initial CLIC stage is optimal for Higgs and
top-quark measurements, allowing for precise determination of the
top-quark mass and Higgs couplings, but also setting stringent constraints on
many BSM scenarios.
Subsequent CLIC stages reaching TeV scales will significantly enhance
precision of SM measurements, and extend indirect BSM searches to
$\mathcal{O}$(100) TeV scales and direct searches for ``new physics''
complementary to those planned at HL-LHC  up to the TeV scales.

\section*{Acknowledgements}

This contribution was supported by the National Science Centre, Poland, the
OPUS project under contract UMO-2017/25/B/ST2/00496 (2018-2021).

\end{document}